\documentclass{article}
\newcommand{\smalllistingsize}{}

\usepackage[backend=bibtex]{biblatex}
\usepackage{dmbiblatex}
\bibliography{verified_static_analysis}

\usepackage{amsmath}
\usepackage{paralist}
\usepackage{graphicx}
\usepackage{txfonts}
\usepackage[title]{appendix}%
\usepackage{xcolor}%
\usepackage{manyfoot}%
\usepackage{listings}%
\usepackage{hyperref}

\lstdefinelanguage{Coq}{
mathescape=true,
texcl=false, 
morekeywords=[1]{Section, Module, End, Require, Import, Export,
  Variable, Variables, Parameter, Parameters, Axiom, Hypothesis,
  Hypotheses, Notation, Local, Tactic, Reserved, Scope, Open, Close,
  Solve, Obligations,
  Delimit, Extract, Inlined, Constant,
  Definition, Let, Ltac, Fixpoint, CoFixpoint,
  Morphism, Relation, Implicit, Arguments, Unset, Contextual,
  Strict, Prenex, Implicits, Inductive, CoInductive, Record,
  Structure, Canonical, Coercion, Context, Class, Global, Instance,
  Program, Infix, Theorem, Lemma, Corollary, Proposition, Fact,
  Remark, Example, Proof, Goal, Save, Qed, Defined, Hint, Resolve,
  Rewrite, View, Search, Show, Print, Printing, All, Eval, Check,
  Projections, inside, outside, Def},
morekeywords=[2]{forall, exists, 
exists2, fun, fix, cofix, struct,
  match, with, end, as, in, return, let, if, is, then, else, for, of,
  nosimpl, when},
morekeywords=[3]{Type, Prop, Set},
morecomment=[s]{(*}{*)},
showstringspaces=false,
morestring=[b]",
morestring=[d],
tabsize=3,
extendedchars=false,
sensitive=true,
breaklines=false,
literate=
    {.}{{.}}0
    {:=}{{$\coloneqq$}}1
    {forall}{{$\forall$}}1
    {exists}{{$\exists$}}1
    {<-}{{$\leftarrow$}}1
    {=>}{{$\Rightarrow$}}2
    {<=}{{$\!\leq\!$}}1
    {>=}{{$\!\geq\!$}}1
    {==}{{\code{==}}}1
    {==>}{{\code{==>}}}1
    {->}{{$\rightarrow$}}1
    {~>}{{$\leadsto$}}1
    {~~>}{{$\leadsto$}}1 
    {<~}{{$\leftsquigarrow$}}1
    {<->}{{$\leftrightarrow\;$}}1
    {\/\\}{{$\wedge$}}1
    {\\\/}{{$\vee$}}1
    {++}{{\code{++}}}1
    {...}{{${\ldots}$}}1
}[keywords,comments,strings]

\lstnewenvironment{lstcoq}{\lstset{language=Coq,basicstyle={\smalllistingsize\tt}}}{}

\newcommand{\true}{\textsf{true}}
\newcommand{\false}{\textsf{false}}

\newcommand{\filename}[1]{\texttt{#1}}
\newcommand{\fileurl}[2]{\href{#1}{\filename{#2}}}
\newcommand{\filen}[2]{\fileurl{https://github.com/AbsInt/CompCert/blob/2ca39a2801d333abcfa3d691620d03abde4e7e37/#1/#2}{#2}}
\newcommand{\filenc}[2]{\fileurl{https://gricad-gitlab.univ-grenoble-alpes.fr/certicompil/Chamois-CompCert/-/blob/7544c6b9a71459799984059bdfb2029ccf992b44/#1/#2}{#2}}

\title{Pragmatics of Formally Verified Yet Efficient Static Analysis, in particular for Formally Verified Compilers}
\author{David Monniaux}

\begin{document}
\maketitle

\begin{abstract}
  Formally verified compilers and formally verified static analyzers are a solution to the problem that certain industries face when they have to demonstrate to authorities that the object code they run truly corresponds to its source code and that it satisfies certain properties.
  
  From a scientific and technological point of view, they are a challenge: not only a number of nontrivial invariants and algorithms must be proved to be correct, but also the implementation must be reasonably effective so that the tools operate within reasonable time.
  Many optimizations in compilers rely on static analysis, and thus a formally verified compiler entails formally verified static analyses.

  In this article, we explain some difficulties, possible solutions, design choices and trade-offs pertaining to verified static analysis, in particular when the solution of the analysis is expressed as some form of tree, map or set.
\end{abstract}

\section{Introduction}
\paragraph{Static Analysis} consists in deriving information about software without actually running it, by analyzing its source or object code. In some cases, static analysis may consist in checking that the program satisfies some stylistic constraints (e.g., not reusing the same name for a global and a local variables), or checking for patterns that often indicate mistakes (e.g., a memory block is freed along the normal exit of a function but not along a side exit).
In this paper, we shall be solely concerned about static analysis that aims at proving that all possible executions of a program satisfy certain properties, most often through some form of \emph{abstract interpretation} \cite{DBLP:journals/logcom/CousotC92}.

Such static analysis may be used for three main purposes:
\begin{inparaenum}[(i)]
  \item ascertaining which areas should be ``manually'' examined by engineers (e.g., if automated analysis can prove that runtime errors are absent from most of the program, the engineers can focus on the remainder);
  \item proving that software behaves correctly, for instance as an argument for authorities in case the software must be qualified for safety-critical applications;
  \item proving that certain conditions for optimizations during compilation are met (e.g., an operand is always nonnegative, so signed extension and unsigned extension coincide on this operand).
\end{inparaenum}
The degree of tolerable uncertainty about the analysis results varies depending on the use. Obviously, it can be a serious issue if static analysis derives incorrect results that result in miscompilation, that is, of the production of object code not matching the semantics of the source code.

\paragraph{Formally Verified Compilation}
Mainstream compilers have bugs \cite{YangCER11,10.1145/2931037.2931074}.
In most industries, bugs caused by compilers are not a major concern compared to the amount of bugs already present in the source code of the application.
In certain safety-critical industries, it is required that the designers of an embedded computing system show that the object code it uses matches the high-level specification, and in particular that the object code matches the source code.
Common approaches to that problem often involved running a well-known compiler with most, if not all, optimizations turned off, so that object code follows source code closely, and some human examinations \cite{DBLP:conf/date/FrancaFLPS11}.
Such a solution is costly both in terms of cost and code efficiency.

Two alternatives approaches have been proposed. One is whole-compilation \emph{translation validation}: the program is compiled with a normal compiler, then a procedure tries to match the source and object codes, perhaps using debugging information, and to prove their correspondence. This however tends to impose some constraints on the compiler and compilation options used, the form of the source code, etc., for the matching heuristics to succeed. To our knowledge, the only large-scale example of this approach is seL4.~\cite{DBLP:conf/pldi/SewellMK13}
The other approach is \emph{formally verified compilation}.
Various tools, based on various formalisms (higher-order logic, Floyd-Hoare style proofs,…) can be used to prove that a program behaves according to a specification. In particular, it is possible to prove that a compiler behaves according to its specification, namely that it compiles programs correctly.
CompCert \cite{DBLP:journals/cacm/Leroy09,DBLP:journals/jar/Leroy09} is such a formally-verified C compiler, used in some safety-critical industries.~\cite{DBLP:conf/date/FrancaFLPS11}
CakeML%
\footnote{\url{https://cakeml.org/}}
is a formally-verified ML compiler.

Compilers need static analysis at certain steps. For instance, they may perform a \emph{points-to analysis} to see if pointers may be aliased---knowing that some pointers may not alias, that is, may not point to the same memory locations, allows certain optimizations, such as swapping a load and a store operations. A formally verified compiler will thus need some form of formally verified static analysis.

Formal proofs come at a significant cost for developers: in development, a correctness proof for a procedure may be significantly larger than the procedure itself and require much more expert work; in maintenance, it must generally be updated whenever the procedure is updated. It is therefore often desirable to minimize the number of properties to prove. One way to achieve this is by splitting the analysis algorithm into an oracle needing no proof, and a formally verified checking procedure (\emph{formally verified defensive programming} \cite{DBLP:books/hal/Boulme21}).

Another issue is efficiency and access to low-level constructs. For instance, \emph{hash-consing} is a well-known approach for speeding up certain symbolic computations, but it requires a global hash table (and possibly auxiliary tables for \emph{memoization} of operations). This global hash table is part of the global state, and thus cannot be easily modeled inside a pure functional language such as Gallina (the language of the Coq proof assistant).

\section{Software Discussed in the Article}
\paragraph{CompCert}
In this paper, we shall discuss some pragmatic choices that have been made in designing the static analyses of the ``official'' releases of CompCert%
\footnote{\url{https://github.com/AbsInt/CompCert}}
as well as the ``Chamois'' branch.%
\footnote{\url{https://gricad-gitlab.univ-grenoble-alpes.fr/certicompil/Chamois-CompCert}}
We however expect our insights to be valid for any kind of formally verified static analysis.

CompCert is a formally verified compiler for a large subset of the C programming language. It is organized into a few unverified frontend steps taking C as input, a formally verified core, then a few unverified steps that produce assembly code.~\cite{DBLP:conf/esop/MonniauxB22} The formally verified core is organized in a succession of passes operating on intermediate languages. Each intermediate language is equipped with a formal operational semantics. Instructions operate over program states, and may optionally emit externally observable events (external function calls, accesses to special CPU registers, accesses to volatile variables…).

Many optimization passes operate over the RTL intermediate representation, which models execution state as a control location inside a current function, an abstract call stack, a memory consisting in memory blocks, and local ``pseudo registers''. A later phase allocates these pseudo registers into stack frames and CPU registers.

The overall correctness theorem of the compiler is that, if compilation succeeds, then the sequence of externally visible events defined in the C source semantics is matched by the sequence of externally visible events defined in the assembly code semantics. (If undefined behavior occurs, then no guarantee is provided.)
This correctness is proved by the composition of simulation proofs for each pass.
Except for some source semantics, all semantics in CompCert are deterministic and the associated simulation proofs are forward simulations.%
\footnote{The semantics of C is nondeterministic: for instance, the compiler is in general free to choose the evaluation ordering of the arguments to operators and function calls \cite[\S6.5.3]{C99}. The simulation proof should thus be backward: any execution of the compiled code should match one of the executions allowed by the source semantics. In addition, the compiler is allowed to assign arbitrary target executions to source executions with undefined behavior; this leads to a rather complex simulation property \cite[\S2.1]{DBLP:journals/cacm/Leroy09}. If, however, the source and target languages are deterministic, this backward property is equivalent to a forward property: if the source program $S$ has a defined behavior $B$, then the target program must also exhibit behavior~$B$.}

The overall correctness proof discusses whole traces of execution. However most proofs arguments are local and deal with the replacement of some number of source steps by some number of target steps.
In the simplest case, simulation is lock-step: one step of the program prior to the transformation is matched by one step of the program after the transformation. A lock-step forward simulation argument is of the form: ``if a source state $s_1$ can take a step to a source state $s_2$, emitting events $e$ (possibly none), and $s'_1$ is a target state that simulates $s_1$, then there must exist a target state $s'_2$ that simulates $s_2$ and so that $s'_1$ can take a step to $s'_2$ emitting the same events $e$'':
\begin{equation*}
  \forall s_1 s_2 s'_1, ~s_1 \rightarrow_e s_2 \land s_1 \sim s'_1
  \implies \exists s'_2,~ s'_1 \rightarrow_e s'_2 \land s_2 \sim s'_2
\end{equation*}

Many transformation or optimization passes rely on the results of a static analysis. For instance, constant propagation on RTL relies on a \emph{value analysis} that establishes that certain pseudo-registers and certain memory locations contain certain values. The simulation argument for the pass then relies on the invariants produced by this analysis being inductive (correct at function entry and correct at the next step if correct at the current step). The state simulation relation $\sim$ refers to these invariants.

Because it is a compiler, CompCert should be reasonably fast, and this is a challenge:
\begin{inparaenum}[(i)]
\item the Coq code handles integers as linked lists of bits, as opposed to machine words, leading to inefficient arithmetic
\item clearly separated transformation passes may be less efficient than passes than perform several operations at once
\item analyses are performed when needed with no provision for preserving their results across passes.
\end{inparaenum}

The \emph{Chamois} branch features additional optimizations and experiments. Since it adds many additional passes, it is even more sensitive to inefficiencies.
The \emph{Verasco} static analyzer (see Section~\ref{sec:verasco}) was implemented on top of CompCert's front-end.

\paragraph{Astrée}
The Astrée static analyzer~\cite{DBLP:conf/birthday/BlanchetCCFMMMR02,DBLP:conf/pldi/BlanchetCCFMMMR03}
verifies that C programs do not reach undefined behaviors, including assertion violations, by automatically deriving inductive invariants.
It originally targeted safety-critical code for avionic control applications.
It is not formally verified, but some of its design choices inspired formally verified tools (Verasco…) and some of the efficiency challenges it faced are found in other tools.

Astrée performs whole program analysis, following structured control flow (with some extra constructs for dealing with \texttt{goto}). It abstracts numerical variables using intervals, octagons and specific abstract domains for control applications (numerical filters).

Because it performs whole program analysis on control programs that typically contain a number of remanent variables%
\footnote{By remanent variables we mean all those that are created at program startup and have indefinite lifetime: global variables, file-local and function-local \texttt{static} variables. A function-local \texttt{static} variable has local scope, but its value is retained from one call to the function to the next, as opposed to an \texttt{auto} (default case) local variable, which is created when coming into scope and destroyed when coming out of scope.}
linear in the size of the program, the performance of the data structures used to map program variables into memory cell indices and memory cell indices into abstract values was very important (\S\ref{sec:non_relational}).

Even though Astrée is not expected to perform as fast as a compiler, industries typically expect it to run (say, during the night) throughout their development process, as some form of continuous integration process, to catch possible problems early.

\section{Tree, Maps and Sets in Static Analyses}
\label{sec:maps_sets}
When implementing static analyses, it is often necessary to use data structures representing maps or sets. We shall briefly see here some of the efficiency challenges they pose, before seeing, in later sections, some of the solutions we brought forth.

\paragraph{Non-Relational Analyses}\label{sec:non_relational}
Static analysis typically maps every control location, say in a procedure, to some information. We consider here the case where this information pertains to the reachable program states at that location.
Such an analysis is deemed \emph{non-relational} if the information is independent across variables; in contrast, a relational analysis will attempt tracking some forms of relationships between variables. A classical example of non-relational analysis is \emph{interval analysis}, which tracks one interval per variable.

Consider for instance the following program:
$y := x; z = x-y$.
We perform interval analysis: to each variable at every location is associated an abstract value that is an interval.
Assume the precondition $x \in [0,1]$, then the analysis will derive $y \in [0,1]$, and then $z \in [-1,1]$, which is correct but a strict over-approximation of the exact postcondition $z = 0$. This postcondition, however, may be reached only by knowing the relationship $x=y$, not just by propagating per-variable information.

Despite that kind of weaknesses, non-relational analyses are extensively used, including inside compilers, because they are quite cheap.
``Official'' releases of CompCert, for instance, have a \emph{value analysis}%
\footnote{See \filen{backend}{ValueDomain.v} and \filen{backend}{ValueAnalysis.v}}
that tracks if a variable is known to be actually a constant, or, if a pointer, whether it points to certain zones. This analysis can, for instance, track that pointers derived from arguments to a function point outside of the stack frame of that function, and thus cannot alias with pointers that are known to point inside the stack frame.
In addition to this value analysis, Chamois has an interval analysis for integer variables~\footnote{See \filenc{backend}{ZIntervalDomain.v}, \filenc{backend}{ZIntervalAnalysis.v} and \filenc{scheduling}{BTL\_ZIntervalAnalysis.v}}, which is used for showing that certain variables are nonnegative (for replacing operations by simpler ones if they operate only on nonnegative numbers) and that certain computations do not overflow 32-bit values and thus can be promoted to 64-bit without changes in semantics.

A common way to implement non-relational analyses is to compute, for every control location, a data structure implementing a map from variable identifiers to abstract values.
Obviously, such data structure should have fast access both for reading and writing values: when an instruction $r:=f(a,b,c,d)$ is analyzed, the analyzer must fetch the abstract values for $a$, $b$, $c$ and $d$ from the structure, apply the abstract operation corresponding to $f$, then write the result to the structure. If the data structure is to be stored for every location, then this write operation should retain the old structure in addition to the new one. In addition, we should avoid needless data duplication, thus old and new structures should share as much as possible.

Obviously, this structure may be implemented as a functional map, through balanced binary trees (as in the Astrée static analyzer \cite[\S6.2]{DBLP:conf/birthday/BlanchetCCFMMMR02}), Patricia trees or similar.%
\footnote{A Patricia tree is a radix tree with radix equal~2.}
In fact, CompCert has successively had two libraries implementing prefix trees mapping positive integers to values through decomposition from low-order to high-order bits.
The second one \cite{DBLP:journals/jar/AppelL23} has the nice property that two extensionally equal maps ($m(k)=m'(k)$ for all $k$) must be actually identical data structures: the maps are canonical, whereas, in the first version, it was possible to have two different data structures representing the same map.
In Coq, it is often easier to work with canonical representations, since this avoids reasoning with respect to an equivalence relation such as extensional equality or semantic equivalence: doing so entails at least tedious proofs that operations behave the same modulo that equivalence, and sometimes one ends up with impossible obstacles, especially if using dependent types.

Unfortunately, such a data structure is inefficient if it must be considered globally, which happens in two cases in static analysis:
\begin{itemize}
\item When two flows of control join at a certain point, such as the end of an if-then-else construct, with maps $m$ and $m'$ then one must construct a map $m \sqcup_{\textit{maps}} m'$ such that $(m \sqcup m')(k) = m(k) \sqcup m'(k)$ where $\sqcup$ is the least upper bound operator for the abstract values.
  This entails going through all keys~$k$.
\item When one checks the invariant for inductiveness, one checks that $m(k) \sqsubseteq m'(k)$ for all~$k$.
\end{itemize}

It was soon recognized when designing the Astrée system \cite[\S6.2]{DBLP:conf/birthday/BlanchetCCFMMMR02} that
if the analysis tracks all variables in the program, including global variables, these global join operations may come to have intolerable cost. These operations have cost linear in the number $|V|$ of variables, but in a program, especially the kind of safety-critical control programs that Astrée or verified compilation targets, the number of global variables is linear in the size $|P|$ of the program. The number of if-then-else operations is also linear in the size $|P|$ of the program. This means that, even for a loop-free program, the total cost of just the $\sqcup_{\textit{maps}}$ operations will grow quadratic in the size of the program, which quickly becomes intolerable.

For CompCert's value analysis, two workarounds are used.
Firstly, the analysis is local to each function, as opposed to Astrée.
Global variables are assumed to contain arbitrary values at function entry, except for read-only variables, which are assumed to contain their initialization value (the map for read-only globals is computed once and for all and thus there are no costs associated to joins).
The analysis tracks changes to global memory inside the function, but the data structure used just has to track a limited number of updates as opposed to tracking the entire memory state.
Secondly, for local variables that are not allocated a memory location (pseudo-registers), only live variables are tracked.
\footnote{See \filen{backend}{ValueAnalysis.v}: information about a variable is cleared when analyzing the instruction at its last use, ``last'' being taken in a total ordering of program locations. This approach is easy to prove sound, since it is always sound to forget information about a variable.}

Another possible approach would be to use a sparse analysis based on single static assignment (SSA) form.~\cite{DBLP:books/hal/Gonnord17}

\paragraph{Data-flow facts}
Many data-flow analyses attach to each control location a set $s$ of dataflow ``facts''.
When several control flows join at a location, the set of facts known to be alway true at that location is the intersection of the set of facts known to be always true at the incoming edges. Therefore, $s \sqcup s'$, the semantic ``least upper bound operation'', is actually $s \cap s'$.%
\footnote{%
This explains why the conventions for lattices in abstract interpretation and dataflow analyses are often opposite, with the ``top'' element in abstract interpretation, meaning ``I know nothing'', being implemented by $\emptyset$, the bottom elements of sets.}

The question then becomes how to implement these sets. The same kinds of tree-like structures used for maps can be used, with values being either $0/1$ or simply always $1$ (because the absence of a key/value association in a map may be interpreted as $0$).
Again, the efficiency problems lie in the global operations: set union, set intersection, and inclusion testing.

\paragraph{Symbolic execution}
One way to validate the results of an optimization phase is to check that the original and the transformed programs are equivalent through symbolic execution.
In the absence of branching control-flow constructs, symbolic execution means executing the program over symbolic inputs, computing intermediate values as terms over these symbolic inputs and the possible arithmetic operations.
If these terms are equal in the outcomes of two programs, then these programs are equivalent.
For instance, $x := 3; y := 3; z := x+1$ and $x := 3; y := x; z := y+1$ are equivalent because both produce $x : 3$, $y : 3$, $z : 3+1$.
Thus checking equivalence of two programs boils down to checking equivalence of terms represented as trees, that is, again, checking that two trees are equal.

Such a system ignores the semantic meaning of operators ($3+1$, $1+3$ and $4$ wil be considered different terms), but it is possible to enrich it by rewriting rules implementing such transformations; again checking equivalence boils down to checking equalities of terms in normal forms with respect to rewriting.
Equivalences of programs containing branching controls, or even loops (through auxiliary invariants), may be checked likewise.
The equivalence requirement may be relaxed to only apply to live variables.

As we shall see in Section~\ref{sec:symbolic_execution}, this symbolic execution approach has been extensively used inside Chamois for implementing optimizations such as basic block or superblock scheduling, loop-invariant code motion and strength reduction of index multiplication in loops.
Efficiency here lies in being able to construct terms, apply rewriting at the root, and check for term equality very efficiently.

\section{Solutions for Efficiency}
\label{sec:efficiency}
If we were implementing a regular compiler (or some other category of symbolic tool, such as a computer algebra system, proof assistant…), a number of implementation ``tricks'' would be available to us to ensure efficiency.
It is however not so easy to use these ``tricks'' in code formalized within a proof assistant, in particular if the proof assistant, such as Coq, views programs as purely functional.

\subsection{Physical Pointer Equality}\label{sec:phys_eq}
The efficiency problem that we pointed out in Section~\ref{sec:maps_sets} can be stated as: when we apply a global operation (inclusion testing, least upper bound\dots) on two maps $m$ and $m'$, the cost of that operation is proportional to the number of variables $|V|$ in the maps even if the maps are very similar.
For instance, at the end of an if-then-else construct
\begin{lstlisting}[language=C]
  if (y < 0) {
    x = 3;
  } else {
    x = 5;
  }
\end{lstlisting}
the interval map will be the same from both branches except for the variables \lstinline|x| and \lstinline|y|, but the least upper bound operation for abstract values will be applied to all variables, even those that have not been touched in either branch.

The solution used in Astrée for least upper bound operations was, when traversing the tree data structures implementing maps from variables to abstract values, to opportunistically detect cases when the subtree that would be produced would be identical to one of the input subtrees, and in this case to return that input subtrees through the same pointer.~\cite[\S6.2]{DBLP:conf/birthday/BlanchetCCFMMMR02}
For instance, if computing the least upper bound $\sqcup_{\textit{maps}}$ of two subtrees given by identical pointers, the procedure would immediately return the same pointer.
The procedure for testing inclusion of two subtrees would first check if the two subtrees were given by identical pointers and return true immediately in that case.
That opportunistic use of identical pointers was key to the efficiency of the analysis.

Can this approach be adapted to a formally verified context? It is tempting to add a predicate $==$, meaning ``physically equal pointers'', and an axiom $\forall x \forall y~ x==y \implies x=y$. Unfortunately, this leads to paradoxes, because $x=y$ means that $x$ can be substituted by $y$ in any context and still yield identical results (``Leibniz equality'').
\begin{figure}{\smalllistingsize
\begin{lstlisting}[language={[Objective]Caml}]
# type t = A of int;;
# let x = A 0 and y = A 0;;
# x = y;;
- : bool = true
# x == x;;
- : bool = true
# x == y;;
- : bool = false
\end{lstlisting}}
\caption{Pointer equality can distinguish between two equal values}
\label{fig:pointer_eq_not_Leibniz}
\end{figure}
Consider the program in Figure~\ref{fig:pointer_eq_not_Leibniz}.
The expressions \lstinline|x == x| and \lstinline|x == y| should yield identical results because $\lstinline|x|=\lstinline|y|$, but they do not.
In short, the problem is that \lstinline|==| allows distinguishing between semantically identical values (here \lstinline|x| and \lstinline|y|), whereas in logic no relation can be finer than equality.

\begin{figure}
\begin{lstcoq}
Axiom tree_phys_eq: tree -> tree -> ?? bool.
Axiom tree_phys_eq_correct: forall t1 t2,
  tree_phys_eq t1 t2 ~~> true -> t1 = t2.
\end{lstcoq}
\caption{Using Boulmé's monad system \cite{DBLP:books/hal/Boulme21,DBLP:conf/vstte/FouilheB14}, a pointer equality (``physical equality'') operator is declared over a tree datatype, inside a ``may return'' monad with Boolean return type (\lstinline[language=Coq]{?? bool}). The axiom states that if this operator has returned true, then the two values are semantically equal.}
\label{fig:phys_eq}
\end{figure}

One possible solution is to model \lstinline|==| as a nondeterministic operation within a nondeterministic ``may return'' monad \cite{DBLP:books/hal/Boulme21,DBLP:conf/vstte/FouilheB14}.
This monad encapsulates possibly non-deterministic computations: $c \leadsto v$ means that the computation $c$ may evaluate to $v$; the difference with an ordinary expression is that it is impossible to derive $v=v'$ from $c \leadsto v$ and $c \leadsto v'$.
With some syntactic sugar, it allows writing Coq programs that use nondeterministic expressions; instead $c : v$ denoting a deterministic computation $c$ of type $v$, we have $c : ??v$ denoting a nondeterministic, possibly nonterminating, computation evaluating to a value of type~$v$.
Here, we consider that physical equality is non-deterministic because it may return different Booleans (Fig.~\ref{fig:pointer_eq_not_Leibniz}) when called twice with parameters that are semantically equal.
We also add an axiom stating that if $x==y \leadsto \true$, then $x=y$ (Coq declarations in Fig.~\ref{fig:phys_eq}): if two pointers are equal then the objects they point to are equal.
Then, the same opportunistic approach as in Astrée could be implemented.

In the code presented in Figure~\ref{fig:phys_eq}, the second \lstinline|Axiom| is one in the logical sense (it states a logical property that will be assumed from then on), while the first just declares a function \lstinline|tree_phys_eq| taking two trees as argument and returning a Boolean in the ``may return'' monad.

If evaluating terms inside Coq itself, \lstinline|phys_eq| will not be available (evaluation stops on axioms, which are considered \emph{uninterpreted functions}). However, CompCert is not directly executed within Coq, but rather the Coq code is extracted to OCaml, which is then compiled and linked together with manually written OCaml code to form the final executable.
It is in particular possible at that point to state that certain Coq axioms, declaring types and functions, are realized by certain OCaml constructs.%
\footnote{See e.g. \filen{extraction}{extraction.v}/\filenc{extraction}{extraction.vexpand}, and \filenc{lib/Impure}{ImpPrelude.v} in Chamois}
When extracting Coq to OCaml, we set up the extraction mechanism so that \lstinline[language=Coq]|?? bool| gets extracted to \lstinline[language={[Objective]Caml}]|bool|%
\footnote{We also set up the extraction mechanism so that, instead of declaring a new OCaml type translating Coq's \lstinline[language=Coq]|bool| type, we reuse OCaml's standard Boolean type. This is standard.},
that is, the monad is elided,
and \lstinline|phys_eq| gets extracted to pointer equality \lstinline[language={[Objective]Caml}]|(==)|.

\paragraph{Shortcut Test}
Another possible formalization, introduced by Jourdan~\cite{DBLP:phd/hal/Jourdan16}, is to declare physical equality as a special ``shortcut'' test that computes a result through a fast path when two terms are known to be equal (through physical equality), under the condition that this fast path returns the same result as the slow path.
The Coq formalization is:
\begin{lstcoq}
 Axiom phys_eq : forall {A B : Type} (x y : A)
   (fast_path slow_path : unit -> B)
   (Hsame : x=y -> fast_path tt = slow_path tt), B.
\end{lstcoq}
This axiom cannot introduce logical inconsistency: it can be implemented naively by just calling the slow path
\begin{lstcoq}
Definition phys_eq_impl {A B: Type} (x y : A)
  (fast_path slow_path: unit -> B)
  (Hsame: x=y -> fast_path tt=slow_path tt):= slow_path tt.
\end{lstcoq}

The following definition extends this scheme to cases when the fast path is correct only if $x=y$:
\begin{lstcoq}
Axiom phys_eq : forall {A B : Type} (x y : A)
  (fast_path : x = y -> B)
  (slow_path : unit -> B)
  (Hsame: forall (eq: x=y), fast_path eq = slow_path tt), B.
\end{lstcoq}

The advantage of this approach is that it does not require reasoning within a monad. However, it imposes some form of local confluence of the computations between the slow and fast paths.

\subsection{Hash Consing}\label{sec:hash_consing}
The opportunistic approach consists in recognizing locally that some value that we are about to construct is equal to a value that we already have (perhaps a parameter to the function), and return that value instead of constructing another occurence of it, so that it the result can be recognized to be identical to that value by pointer equality.

A more general approach is \emph{hash consing} \cite{DBLP:conf/ml/FilliatreC06}: all values created so far for a particular datatype are stored in a \emph{hash} table, and, when a value is about to be \emph{cons}tructed,%
\footnote{While the name \emph{hash consing} is associated with Lisp terminology from the 1970s, the idea of hash tables and hash-consing appeared as early as 1958 in the Soviet Union, in the context of compilation. \cite{10.1145/368892.368907}}
the table is checked for an existing copy of that value, which is returned instead if it exists. This ensures that no two copies of the same value can coexist (at different memory locations) in the system, and thus pointer equality is equivalent to value equality.

Such a hash table would continue growing and storing useless values. One possible workaround is to make the hash table local to a phase of the computations and discard it at the end of the phase. Another is to replace the hash table by a \emph{weak hash table},
\footnote{In particular, functor \fileurl{https://ocaml.org/manual/5.1/api/Weak.Make.html}{Weak.Make} in OCaml. Weak hash tables are however available in other languages, e.g. \fileurl{https://docs.oracle.com/javase/8/docs/api/java/util/WeakHashMap.html}{WeakHashMap} in Java.}
so that values considered unreachable by the garbage collector of the execution platform are removed from the hash table (a value being reachable from the weak hash table does not make it considered as reachable by the garbage collector, as opposed to a normal hash table).

Because of the importance of hash-consing for implementing certain forms of symbolic computations, there has been some interest in how to use it from formally verified software written in a purely functional language, in particular with the Coq proof assistant~\cite{DBLP:journals/jar/BraibantJM14}. We shall now discuss various difficulties, workarounds, and trade-offs involved in this.

\paragraph{As an untrusted oracle}
One possible implementation of hash consing is to use the hash table as an untrusted oracle. When we are about to construct a term $t=C(x, y, z)$ where $C$ is a term constructor and $x,y,z$ are subterms, we query the hash table for a copy $t'$ of that term. We do not trust this $t'$: we check that it is of the form $C(x',y',z')$, and then retain it if $x==x'$, $y==y'$ and $z==z'$ all $\leadsto \true$, which should always be the case if the hash table works properly.

A weakness of that system is that, even though it will never create two identical copies of the same term (provided all term constructions go through the process described above), and thus pointer equality is equivalent to equality for practical purposes, pointer equality is not \emph{provably} equivalent to equality.
In other words, we cannot conclude from the fact that $x==y \leadsto \false$ that $x \neq y$, even though this works in reality.
Furthermore, this system runs the whole computation inside a ``may return'' monad, which complicates proofs and forces the whole of the program to be executed inside that monad.

Because of the inconvenience of rewriting the whole of CompCert inside a monad, certain optimizations in Chamois, which are validated using a symbolic execution engine based on hash-consing \cite{DBLP:journals/pacmpl/SixBM20}, use an ``unsafe exit'' from the ``may return'' monad they were using. This ``unsafe exit'' turns a nondeterministic reduction $e \leadsto v$ into an ordinary value~$v$.
The reason why this is unsafe is that if we apply it to $e$ such that $e$ may return different values $v \neq v'$ ($e \leadsto v$ and $e \leadsto v'$), then considering this return value as deterministic leads to $v = v'$ and then an absurd case.
The designers of Chamois however considered that this was not an issue, since this would somehow involve a case where the same optimization phase would be deliberately run twice on the same input, then the (nondeterministic) outputs compared and an absurd case entered if they differ. 




\paragraph{Efficiency Tradeoff}
\begin{figure}
\begin{lstcoq}
Inductive tree :=
| Node : tree -> tree -> tree
| Leaf : nat -> tree.

Fixpoint tree_eqb (t t' : tree) :=
  match t, t' with
  | (Leaf n), (Leaf n') => Nat.eqb n n'
  | (Node l r), (Node l' r') =>
      (tree_eqb l l') && (tree_eqb r r')
  | _, _ => false
  end.

Fixpoint tree_eqb_fast (t t' : tree) : ??bool :=
  DO cmp <~ tree_phys_eq t t';;
  if cmp then RET true else
  match t, t' with
  | (Leaf n), (Leaf n') => RET (Nat.eqb n n')
  | (Node l r), (Node l' r') =>
      DO cmp_l <~ tree_eqb_fast l l';;
      if cmp_l then tree_eqb_fast r r'
      else RET false
  | _, _ => RET false
  end.

Lemma tree_eqb_fast_correct: forall t1 t2,
    WHEN tree_eqb_fast t1 t2 ~> b THEN b = tree_eqb t1 t2.
\end{lstcoq}
\caption{A simple tree datatype with a naive equality test and a fast ``shortcut'' equality test, using the pointer equality defined in Fig.~\ref{fig:phys_eq}. It is possible to prove that the two coincide (below) and that they implement equality testing.}
\label{fig:tree_eqb}
\end{figure}

Let us compare now the naive and the shortcut equality tests in a context where trees are always produced by hash-consing, and thus there exist no two identical subtrees at distinct addresses.
When given two identical trees, the naive test will traverse them fully, with complexity linear in the size of the tree. The shortcut test will terminate immediately with a positive answer.
When given two different trees $t$ and $t'$, the naive test will still need to fully traverse identical subtrees, until it finds a path from the root that leads to different items in $t$ and~$t'$. This path forms a \emph{witness} that the two trees are different.%
\footnote{By \emph{witness} we mean a piece of data that is sufficient to establish the property. Here, to establish that $t$ and $t'$ are different, it is sufficient to exhibit a path that leads to different nodes in the two trees.}
The shortcut test will avoid these traversals and instead converge directly on such a witness.
Its complexity is thus bounded by the minimum of the depths of $t$ and~$t'$.

Arguably, if trees are hash-consed, it should not be necessary to find a witness path for the difference of two trees, because the pointer equality test at the root gives the answer. However, in order to conclude that if the pointer equality test yields false, then the trees are different, we must use the invariant that all trees are created by hash-consing, a very strong assumption that cannot be directly expressed within the system. If we do not have this assumption and just the assumption that pointer equality (through a ``may return'' monad) implies tree equality, we end up with a less efficient equality test in case the trees differ.

\paragraph{Inequality as error}
In Chamois, phases validated by symbolic execution check equalities of terms by pointer equality, and the program exits immediately if it returns false. There is therefore no need to generate a witness that two terms are not equal.

\paragraph{As a trusted oracle}
In the ``untrusted oracle'' hash consing system, nothing prevents the application code from creating trees not going through the hash consing system. If we want the property that all trees go through hash-consing, then an approach is to ask Coq to extract the datatype to a specific OCaml type, with user-specified OCaml constructor and ``match'' operations. Then, whenever Coq code creates a tree in that datatype, it will call the ``smart constructor'', which will perform hash consing.~\cite[\S6.1]{DBLP:journals/jar/BraibantJM14}
Such an approach also makes it easy to hide fields, such as unique identifiers or hash values, that are not relevant at the Coq level.

This amounts to trusting the workings of the hash table and the hashing mechanism (in particular, that we will always be able to find extant elements in the table). If the hash table is weak, this means we trust its non-trivial interaction with the garbage collector of the execution platform.
This extends the \emph{trusted computing base} of the static analyzer or compiler.
This is the choice that was made for the ``hashed sets'' library for sets of positive numbers, used to represent sets of dataflow facts in the CSE3 global common subexpression elimination phase of Chamois \cite{DBLP:journals/tecs/MonniauxS23}.

The argument here is that CompCert's trusted computing base \cite{DBLP:conf/esop/MonniauxB22} already includes Coq itself, which uses OCaml hash tables internally, so it does not seem that trusting the same hash table in extracted code adds much to it.

\paragraph{Using a Hash-Consed Backend Language}
An alternative to using a custom constructor would be to use a special backend language with automatic hash-consing of datatypes, an approached pioneered by HLISP~\cite{Goto_HLISP}.
The GimML language,%
\footnote{\url{https://projects.lsv.fr/agreg/?page_id=258} Formerly HimML.}
from the ML family~\cite{GimML_refman,Goubault94himml:standard,JG:Sharing}, automatically performs hash-consing on datatypes on which it is safe to do so, which is for instance used to implement efficient finite sets and maps.

\section{Invariant inference}
Static analysis of programs containing loops must often compute inductive invariants. These invariants are often obtained by an iterative fixed-point (or post-fixed-point) computation.

\subsection{Fixed-Point Computation}
\paragraph{Iterative Computation}
Invariants are obtained as fixed points of certain operators.
Consider transition systems where transitions are of the form $(p,\sigma) \rightarrow (p',\sigma')$, where $p,p' \in P$ are control locations and $\sigma,\sigma' \in \Sigma$ are data states.
An \emph{invariant} over that transition system is thus a mapping from $P$ to the powerset of $\Sigma$, associating to each control location a set of states that must contain the states reachable at this location.

A state $(p,\sigma)$ is reachable if and only if it is either an initial state (typically, there is a $p_0$ initial location with an associated set $\Sigma_0$ of initial data states), or if there exists a reachable state $(p_{\textit{pre}},\sigma_{\textit{pre}})$ such that  $(p_{\textit{pre}},\sigma_{\textit{pre}}) \rightarrow (p,\sigma)$.
An \emph{inductive invariant} $I$ is thus a mapping from $P$ to the powerset of $\Sigma$ such that
\begin{inparaenum}[(i)]
\item it contains initial states ($\Sigma_0 \subseteq I(p_0)$)
\item it is inductive: for all $(p,\sigma) \rightarrow (p',\sigma')$ with $\sigma \in I(p)$, then $\sigma' \in I(p')$.
\end{inparaenum}
An inductive invariant is an invariant, but an invariant may be noninductive.

Assume we have an abstract domain $\Sigma^\sharp$ for representing subsets of $\Sigma$. Assume also that we have a function $S$ such that, for $p \in P$ and $\sigma^\sharp \in \Sigma^\sharp$, $(p,\sigma^\sharp)$ is a finite set of pairs $(p',{\sigma^\sharp}')$ such that if $(p,\sigma) \rightarrow (p',\sigma')$, $\sigma \in \gamma(I^\sharp(p))$, then there exist $(p',{\sigma^\sharp}') \in S(p,I^\sharp(p))$ such that $\sigma' \in \gamma({\sigma^\sharp}')$.
In other words, this function $S$ associates to each control location a finite set of $(\textit{successor},\textit{abstract state})$ pairs; most instructions have only one successor, branching instructions have several.
In the simplest cases, the successor abstract state will be the same regardless of the successor, but it may be useful to have differing abstract states, for instance to reflect the information brought in by the condition on the branching instruction (for instance, after a condition $i=42$ we know that $i=42$).
An inductive invariant $I^\sharp$ for the transition system in the abstract domain is a mapping from $P$ to $\Sigma^\sharp$ such that
\begin{inparaenum}[(i)]
\item it contains an abstract version of the initial states ($\sigma_0^\sharp \sqsubseteq I^\sharp(p_0)$)
\item it is inductive: for all $(p',{\sigma^\sharp}') \in S(p,I^\sharp(p))$, then $ {\sigma^\sharp}' \sqsubseteq I^\sharp(p')$.
\end{inparaenum}
These properties are decidable by a simple procedure, assuming $\sqsubseteq$ and $S$ are computable.

The usual approach to solving such a problem is a workset-based algorithm, in which the workset contains a list of states whose successors may not contain yet the elements that are propagated. The initial workset $W$ contains just $p_0$, and $I^\sharp(p_0)$ is initialized to $\sigma_0^\sharp$.
As long as $W$ is nonempty, a $p$ is picked from it, and for every $(p',{\sigma^\sharp}') \in S(p,I^\sharp(p))$, the algorithm checks if
${\sigma^\sharp}' \sqsubseteq I^\sharp(p')$; if not, $I^\sharp(p')$ is replaced by its ``least upper bound'' with ${\sigma^\sharp}'$, and $p'$ is added to~$W$.

The ``least upper bound'' operator does not actually need to be the least upper bound, it just has to yield a result greater than its operands. In lattices with infinite ascending chains, one usually replaces it with a \emph{widening} operator, which ensures converges in a finite number of iterations.
The widening operator may be applied only at a selected subset of control locations sufficient to break all cycles in the control-flow graph.

If this algorithm terminates, and thus reaches a \emph{fixed point}, then this fixed point is an inductive invariant.
Under some monotonicity condition for $S$%
\footnote{Note that the value analysis in CompCert is not monotone and that the ``least upper bound'' operator is not the least upper bound. See \href{https://github.com/AbsInt/CompCert/pull/490}{issue 490}.}
and with the assumption that the least upper bound operator is really the least upper bound, it will compute the least inductive invariant in the abstract domain.

The order in which elements are picked from $W$ is unimportant for correctness, but is important for efficiency.
For instance, if a procedure consists in two successive loops (or, in terms of graphs: its control-flow graph consists of two strongly connected components), it is more efficient to first compute the fixed point of the first loop, then that of the second loop, rather than the two at the same time.
This is achieved by sorting control locations in reverse postorder and picking the least element from $W$ with respect to this ordering. $W$ can be implemented as a heap.%
\footnote{In CompCert, \filen{backend}{Renumber.v} renumbers the control locations of a procedure so that the entrypoint is maximal, and one picks the maximal element in the workset.}

This algorithm, also known as Kildall's algorithm, is implemented in the official releases of CompCert, with the restriction that all edges outgoing from the same control location receive the same abstract state. Chamois also has another fixpoint algorithm lifting that restriction.
In neither case, the convergence of the algorithm is proved: the algorithm, unless $W$ becomes empty, iterates up to a very large ``fuel'' natural integer, and if it reaches it, gives up and returns an error.
Proving convergence would entail arguing about the absence of infinite ascending chains in $X^\sharp$ and the finiteness of~$P$.

It is not a problem in practice that the analysis can report an error; in this case one can either give up on the optimization that requested the analysis, or safely use $\top$ (``anything is possible'') at all locations.

\paragraph{Fixed-Point Checking}
The above approach directly computes an inductive invariant in a formally verified manner.
Another way is to compute the invariant using an oracle, and then check that it is inductive using a verified procedure.

The elements of the static analysis lattice are likely to be maps (in the case of a non-relational domain, mapping variables to abstract values) or sets (in the case of dataflow analysis). Checking that a fixed point is reached amounts to inclusion testing or even, depending on how the fixed point problem is formulated, to equality testing.
For efficiency, one may want to apply the methods discussed in Section~\ref{sec:efficiency}: inclusion tests apply ``shortcuts'' when identical subtrees (thus identical submaps or subsets) are detected, and hash-consing ensures that identical subtrees actually get identical pointers.

\subsection{Data-Flow Facts}
In data-flow analysis, the lattice $\Sigma^\sharp$ of abstract elements is the powerset of a finite set $F$ of elementary dataflow facts.
One difficulty is that the set $F$ may not be known in advance. In fact, it may be advantageous to dynamically enrich $F$ during the fixed point computation. When an elementary fact is to be used, say $x = y+z$, it is looked up in a hash table that associates an integer to it; if it does not exist in the table, a fresh index is associated to it.
We may also compute auxiliary tables, such as, for every variable $v$, the set of dataflow facts that are to be invalidated by a write to $v$ (e.g. the fact $x=y+z$ is to be invalidated by writes to $x$, $y$ or $z$).

At the end of the fixed point computation, we thus have, among other information
\begin{inparaenum}[(i)]
  \item a mapping from $P$ to subsets of the final $F$, represented as sets of integer indices
  \item a table mapping these indices to their semantics as elementary dataflow facts, which was being updated during the fixed point computation but which can be now taken as a read-only data structure.
\end{inparaenum}
Note that we may have other tables, but since we have not proved that they are updated consistently, we cannot easily use them in subsequent verified computations.
We thus rebuild auxiliary tables in a verified manner, if necessary.

We then check that the computed fixed point is truly inductive, in a verified manner. This is how the data-flow facts used for global common subexpression and condition elimination (CSE3 pass) are established in Chamois~\cite{DBLP:journals/tecs/MonniauxS23}.

In Chamois, the integer indices associated to data-flow facts are positive, and the sets of positive integers are represented as binary trees indexed by the binary decomposition of the integers.
These binary trees constitute a canonical representation: a set may be represented only by one tree, and thus semantics equality is equivalent to structural equality.
These binary trees are built using hash-consing, and thus structural equality is equivalent to pointer equality.
Many operations on the sets (equality testing, union, intersection, inclusion testing, …) are sped up by checking for shortcuts when some subtrees are equal, whch boils down to pointer equality.

Chamois uses hash-consing as a trusted oracle when it comes to binary trees representing data-flow facts: the constructor and pattern-matching operations are replaced through the extraction mechanism by suitable OCaml code, and the equality test is mapped to pointer equality.%
\footnote{\filenc{lib}{HashedSet.v}}
This choice may be considered excessive; it would have been sufficient to just use pointer equality to short-cut inclusion tests, just using as an axiom that pointer equality implies structural equality. However, this would have entailed programming inside a ``may return'' monad, as though the algorithms were nondeterministic, even in cases where it can be proved that the result is deterministic (inclusion testing has a uniquely defined Boolean return value).
Stronger attention was awarded to the ease of expressing results simply than to maximal reduction of the trusted computing base.

\section{Symbolic Execution}
\label{sec:symbolic_execution}
Another kind of static analysis used for proving the correctness of optimization phases in Chamois is \emph{symbolic execution} \cite{DBLP:journals/pacmpl/SixBM20,Gourdin_AFADL21,Gourdin_et_al_OOPSLA23}.
The basic idea is that two sequences of instructions are equivalent if and only if they leave the same final results in the variables, regardless of the order in which they did the operations. An extension of this idea is to consider only those variables that are live at the end of the computation.%
\footnote{In the case of programs with operations that may trap, such memory accesses (in case of invalid addresses) or division (if division by zero is trapping), there is also the requirement that a program may be transformed into another only if the set of expressions that may trap in the second program is included in that for the first.}
This equivalence can be established by computing, for every live variable at the end of the computation, a term expressing it as a function of the variables at the beginning of the computation. These terms are obtained by applying the operations in both programs symbolically.
This is for instance used to show that the code after scheduling performs as the one before scheduling.

For instance, if $x$ and $v$ are not live at the end of the computation, these two programs are equivalent:\\
\parbox{0.45\columnwidth}{
$u := x+y$\\
$z := x+y$\\
$t := x-y$\\
$v := x-y$}
\hfill
\parbox{0.45\columnwidth}{
$u := x+y$\\
$t := x-y$\\
$x := 0$\\
$z := u$}\\
and this can be established by computing symbolic forms: at the end of the computation, in both cases, $y = y_0$, $z = x_0+y_0$, $t = x_0-y_0$, $u = x_0+y_0$ where $v_0$ denotes the initial value of variable~$v$.
Note that, as seen on the second program when performing $z := u$, doing this symbolic execution naively may duplicate expressions. Ideally, we would like $x_0+y_0$ to be stored only once, and only a pointer to it be copied; thus terms should form a DAG (directed acyclic graph), not individual trees.

After symbolic execution has been done, we need to check that for every live variable, the final symbolic terms are identical. In practice, this will always be the case, unless there has been some bug in the optimization phase. We thus need to optimize the case where the terms are equal.
If we check term equivalence naively, by traversal, the complexity of the check will be linear in the size of the terms as trees. That size can be exponential in the length of the programs to be analyzed, as in the following example:
$a_1 := \textit{op}(a_0,a_0)$;
$\dots$;
$a_n := \textit{op}(a_{n-1},a_{n-1})$
where $a_n$ is a complete binary tree of depth~$n$.
In contrast, the size of $a_n$ as a directed acyclic graph (DAG) is linear in~$n$.

The solution for this is, again, hash-consing. The terms are hash-consed, and the optimization is accepted only if the final terms are equal in the sense of pointer equality. Note that the correctness of that approach only relies on pointer equality implying term equality. In particular, it does not require pointer disequality implying term disequality: should this condition not be met (perhaps due to a bug in hashing), the only risk would be that a correct optimization result would be refused, with the compiler skipping that optimization or terminating with an internal error. Such problems can be weeded out by careful testing~\cite{DBLP:conf/tap/MonniauxGBL23}.

The symbolic execution system in Chamois is implemented inside a ``may return'' monad. The hash tables and monad are discarded at the end of the optimization phase. Again, this is theoretically ``unsafe'', but the only way this could create an issue is if the same optimization was run twice and the results compared so as to lead to an absurd case in case they differed.

Symbolic execution and expressions being tested for purely syntactic equivalence have limitations. For instance, the trees representing the integer expressions $a+(b+c)$ and $(a+b)+c$ are different, but they are semantically equivalent.
For certain optimizations, such as \emph{strength reduction}, it is necessary to identify some syntactically different expressions; this can be achieved by applying suitable rewriting rules along with the symbolic computation, so as to compare canonical forms at the end.~\cite{Gourdin_et_al_OOPSLA23}.

Furthermore, both loop-invariant code motion and strength reduction need invariants (e.g. ``variable $t$ in the transformed program stands for expression $p+8 \times i$ in the original program''). These invariants are computed by untrusted oracles, and are checked for correctness by the symbolic execution engine using the rewriting rules.~\cite{Gourdin_et_al_OOPSLA23}

\section{Related Work: Other Forms of Static Analysis}
We have so far discussed formally verified static analysis from the point of view of analyses used for optimizing compilation. Let us briefly discuss some other forms of analyses that are commonly used for program verification.

\subsection{Relational Abstract Domains: Convex Polyhedra}
\label{sec:verasco}
The Verasco project%
\footnote{\url{http://compcert.inria.fr/verasco/}} \cite{DBLP:phd/hal/Jourdan16,DBLP:conf/popl/JourdanLBLP15}
aimed at fitting CompCert with a formally verified static analyzer capable of automatically proving certain properties, such as the absence of certain runtime errors (buffer overflows, arithmetic overflows…).%
\footnote{%
  In other words, that project aimed at implementing a formally verified, simpler analogue of tools such as Astrée \cite{DBLP:conf/birthday/BlanchetCCFMMMR02,DBLP:conf/pldi/BlanchetCCFMMMR03} or Frama-C value analysis.}
It could use both non-relational (intervals) and relational (convex polyhedra) numeric analyses.

Verasco uses data structures with sharing much like Astrée, implemented through physical pointer equality~\cite[Ch.~9]{DBLP:phd/hal/Jourdan16}.
The convex polyhedra libary VPL%
\footnote{\url{https://github.com/VERIMAG-Polyhedra/VPL}}
\cite{DBLP:conf/synasc/BoulmeMMPY18,DBLP:phd/hal/Marechal17,DBLP:conf/sas/MarechalMP17,DBLP:conf/vmcai/MarechalP17,DBLP:conf/itp/BoulmeM15,DBLP:phd/hal/Fouilhe15,DBLP:conf/vstte/FouilheB14}, despite being based on a constraint-only representation that supposedly scales better than the conventional double representation (generators and constraints) with respect to dimension, also had scaling issues, as expected from such a highly relational analysis.
The relational analyses would likely have scaled better if applied to select subsets of the variables (``packs''), as done in Astrée.~\cite{DBLP:journals/lisp/Mine06}

The design choices of the VPL reflect some of the concerns and insights described in this paper: inclusion or equality tests should be fast, and it is often easier to implement a verified operation as the composition of an unverified oracle and a verified checker.

The conventional approach to polyhedral computations is through ``double representation'', where a polyhedron is represented both as a system of \emph{constraints} (faces) and a system of \emph{generators} (vertices, and, in the case of unbounded polyhedra, rays and lines). The two representations are dual, and it is possible to move from one to the other using various algorithms. One representation may also be used to eliminate redundancies from the other. The problem with this approach, from the point of view of a verified analyzer, is that it is unsound to omit generators, and thus the conversion from constraints to generators must be shown not to skip any. VPL eschewed this approach and instead represented polyhedra by constraints only.

Operations on constraints defining polyhedra may be justified by, in essence, showing that certain constraints are logical consequences of others.
This entailment may be justified by showing that the consequence is a combination of the antecedent constraints with nonnegative coefficients, often known as \emph{Farkas coefficients}.%
\footnote{Farkas' lemma states that a linear inequality is a consequence of other linear equalities if and only if it can be expressed as a nonnegative combination of these (the result extends to affine inequalities by allowing relaxation of the constant coefficient). The ``if'' part is trivial but the ``only if'' part is a bit more involved, it may be for instance established by instrumenting Fourier-Motzkin elimination. Farkas' lemma is closely related to the strong duality theorem in linear programming.} 
For instance, if we want to show that the projection on $x$, parallel to $y$, of the polyhedron defined by $x+y \leq 1$ and $x - y \leq 2$  is included in the $x \leq 3/2$, then it is sufficient to recognize that by multiplying $x + y \leq 1$ by $1/2$ and $x - y \leq 2$ by $1/2$ and adding them, we obtain $x \leq 3/2$.
A procedure computing the projection of a polyhedron can thus justify that its result includes the correct projection by providing, for each constraint it outputs, a Farkas certificate establishing that it is a consequence of the constraints in the original polyhedron.
This generalizes to other operation than projections: Farkas certificates prove that the result polyhedron includes the polyhedron to be computed.
This inclusion is sufficient for proving soundness
\footnote{These certificates are however insufficient for proving that the computed polyhedron is exactly the projected polyhedron sought, but that is not needed for soundness of the analysis.}

In the original design of the VPL, certificates were computed explicitly and fed to the verifier, which entailed much bookkeeping. The system was later redesigned so that certificates are not explicitly computed, but rather appear as elements of a data-type that can only be manipulated through certain basic operations that are shown to preserve the property that they are valid certificates.~\cite{DBLP:books/hal/Boulme21}
This simplifies the design (the number of lines of code was halved) and slightly improves performance.

Despite the difficulties involved in the double representation approach, some recent work proposes formally verified yet effective ways of computing the edge-vertex graphs~\cite{DBLP:conf/cpp/AllamigeonCS23}. Again, that approach uses certificates that certain properties are correct (these certificates are later checked for correctness by formally-verified code) as opposed to proving total correctness.

\subsection{SAT / SMT Solvers}
SAT and SMT solvers have been extensively used to check properties of hardware and software systems, and even to prove or disprove mathematical conjectures~\cite{DBLP:conf/sat/HeuleKM16}. When a SAT solver answers positively and provides a purported model, it is easy to check if it is truly a model. When a SAT solver answers negatively, there is no such obvious witness. SAT solvers may however be instrumented to produce some kind of trace of their execution, which can then be verified independently, a feature now considered \emph{de rigueur}.\footnote{%
  In fact, because of the general unreliability of negative results from SAT solvers unless such precaution is taken, certificates of unsatisfiability have been required for UNSAT tracks in the SAT competition since 2013.}
The scientific challenge is to design a certificate format that is both compact (full trace logging is too expensive) and yet relatively easy to check (so that the verifier can be kept simple). Some tools may be used to reprocess and simplify the certificate.

It is possible to write a formally verified checker for SAT with checking time on the same order as the solving and proof processing times~\cite{DBLP:books/hal/Boulme21}.
The SMTCoq tactic takes it further: it calls external SMT solvers and processes the certificates they produce to reach a goal in Coq~\cite{DBLP:conf/cav/EkiciMTKKRB17,DBLP:phd/hal/Keller13,DBLP:conf/cpp/ArmandFGKTW11}.
Again, the same principle applies: an untrusted solver is coupled with a verified checker.

SAT/SMT solvers cannot normally be used to analyze programs containing loops, except by unfolding them to a finite depth. However, there exist some approaches (IC3/PDR in particular) that use SAT/SMT as subprocedures and are capable of proving safety properties.
The reliability of these tools was once not too great \cite[\S6]{DBLP:conf/sas/MonniauxG16}. Producing efficient yet formally verified versions for them seems a challenge.

\section{Future Work}
The various approaches that we described to implement hash-consing had various drawbacks.
A solution would be to use a state monad directly implemented in Coq, as opposed to using extraction tricks, and carry the state of the hash table inside that monad.
This entails implementing the hash table in Coq.
In older versions of Coq, the array in which the hash table is stored had to be itself implemented using some kind of functional map structure (ordered tree, binary tree\dots), which was inefficient.
Thus, hash-consing implemented in this fashion was more of an academic exercise than for actually running it~\cite[\S5]{DBLP:journals/jar/BraibantJM14}, though some have successfully pushed to approach to a full ROBDD implementation complete with a stop-and-copy garbage collector.~\cite{DBLP:conf/cpp/ChavanonBN24}

A more modern approach would use Coq's relatively recent support of native integers and native arrays \cite{DBLP:conf/itp/ArmandGST10}\cite[\S2.1.13, ``Primitive objects'']{Coq_manual}.
The native integers would be used to implement the hash functions efficiently, and the native arrays would be used for the table itself.
These arrays are \emph{persistent}: the store operation is defined to return a new version of the array; old versions of the array can still be accessed, but the implementation is optimized for the case where only the most recently updated version of the array is accessed. Internally, only the last version of the persistent array is retained, inside a regular (mutable) array, and the previous versions are stored as explicit deltas from this last version.~\cite[\S2.3]{DBLP:conf/ml/ConchonF07}
If the previous versions are no longer used, these deltas will eventually get garbage-collected.
There exist extraction configurations so that, when extracting Coq to OCaml, these persistent arrays get extracted to an OCaml implementation along these lines. After extraction, we would thus obtain a hash table similar to one we could have implemented manually in OCaml.

It remains to be seen how exactly to use native integers and arrays to implement an efficient hash table for hash-consing, reflecting the necessary invariants (all extant objects have been allocated in the table).
It seems a much more difficult task to make this table ``weak'', since this amounts to incorporating part of garbage collection (if only some system of reference counting) inside the formalization, whereas normally garbage collection is left to the runtime system.
Perhaps, again, it is best to renounce collecting the terms that have been created and left unused during the use of the hash table, and instead wait until the state monad is exited and the hash table is discarded.
This fits the use of hash consing inside an optimization or code transformation phase, where everything needed for the internals of the phase can be discarded at the end of the application of that phase (perhaps even at the end of the application of that phase to a particular function).

Exiting the may-return monad abruptly at the end of an optimization phase is most likely not dangerous, but is inelegant. A better approach would be to allow exiting the may-return monad if it always returns the same value as some deterministic computation.

\section{Conclusion}
We have successfully implemented and formally verified a number of optimizations (prepass and postpass scheduling, loop unrollings and rotations, global common subexpression elimination, loop-invariant code motion, store motion, strength reduction…) that were not present in the official releases of CompCert~\cite{DBLP:phd/hal/Six21,DBLP:journals/tecs/MonniauxS23,DBLP:journals/pacmpl/SixBM20,Gourdin_et_al_OOPSLA23,Gourdin_AFADL21}.
Most of these transformations involve some form of static analysis to establish their correctness, whether this static analysis computes some invariants or establishes the equivalence of two blocks by symbolic execution.

While these optimizations may appear to be well-known, and are generally available in mainstream, unverified compilers such as GCC or LLVM, there was on every occasion significant work to be done for identifying the necessary invariants, the necessary properties to be proved, for distinguishing what actually needed to be formally verified from what was not, and for formalizing the optimization in a way that made proofs tractable.
We echo here remarks often made about the formalization of semantics, algorithms, or mathematical proofs:%
\footnote{For instance in talks given by Georges Gonthier about his work on the four-color theorem and the Feit-Thomson theorem.}
badly chosen formalism often allows small-scale works and it is only when attempting larger proofs or algorithms that one will encounter difficulties.
We will even go as far as to say that formalizing known optimizations helps understand them better.~\cite{Avigad_AMS}

There is often a choice to be made between proving \emph{complete correctness} of the analysis (the analysis always succeeds and compute a correct result) and \emph{partial correctness} (if the analysis succeeds, then it computes a correct result). The latter is often much easier to prove than the former: there is no need to prove termination (one may either have part of the procedure in untrusted code with no termination requirement, or use a high maximum number of iterations and fail if it is reached), and one can split the analysis into an untrusted oracle and a formally verified checker.

Even at the level of the individual test, we can examine closely what is actually needed or not. For instance, we may check whether two structures are equal (``are these two maps equals'' when searching for an invariant, ``are these two terms equal'' in symbolic execution), but only the positive answer needs to be correct. If a negative answer is produced instead of a positive, the only consequences would be that needless iterations of a search would be run, or an optimization would be refused whereas it could have gone through.

One key insight is that we can apply different standards of proof to different properties that we expect of an algorithm or analysis scheme.
We may, for instance, formally verify soundness, but prove optimality only on paper, and establish performance by experimental measurements.~\cite{DBLP:conf/tap/MonniauxGBL23}

\section*{Acknowledgments}
We wish to thank Sylvain Boulmé for his insights and comments.

\sloppy
\printbibliography
\end{document}